\documentclass [6pt,a4paper]{article}
\usepackage{bbm}
\usepackage{amsfonts}
\usepackage{mathrsfs}
\usepackage {amssymb}
\usepackage {amsmath}
\usepackage{amsthm}
\usepackage{latexsym}
\def\dse#1{\vskip 0.6cm\noindent
        {\large\bf #1}
        \vskip 0.4cm}

\def\dse#1{\vskip 0.6cm\noindent
        {\large\bf #1}
        \vskip 0.4cm}
\usepackage{lastpage}
\usepackage{epsfig}

\begin{document}
\begin{center}
\textbf{\large{On $\mathbb{Z}_{2}\mathbb{Z}_{2}[u]$-$(1+u)$-additive constacyclic}}\footnote { E-mail
addresses:
lpmath.126@.com(P.Li), DW932524047@163.com(W.Dai), kxs6@sina.com(X.Kai).}\\
\end{center}

\begin{center}
{ { Ping Li, \  Wei Dai,\  Xiaoshan Kai} }
\end{center}

\begin{center}
\textit{\footnotesize School of Mathematics, Hefei University of
Technology, Hefei 230009, Anhui, P.R.China} \\
\end{center}

\noindent\textbf{Abstract:} In this paper, we study $\mathbb{Z}_{2}\mathbb{Z}_{2}[u]$-$(1+u)$-additive constacyclic code of arbitrary length. Firstly, we study the algebraic structure of this family of codes and a set of generator polynomials for this family as a $(\mathbb{Z}_{2}+u\mathbb{Z}_{2})[x]$-submodule of the ring $R_{\alpha,\beta}$. Secondly, we give the minimal generating sets of this family codes, and we determine the relationship of generators between the $\mathbb{Z}_{2}\mathbb{Z}_{2}[u]$-$(1+u)$-additive constacyclic codes and its dual and give the parameters in terms of the degrees of the  generator polynomials of the code. Lastly, we also study $\mathbb{Z}_{2}\mathbb{Z}_{2}[u]$-$(1+u)$-additive constacyclic code  in terms of the Gray images.\\


\dse{1~~Introduction}
Cyclic and constacyclic codes have considered to be one of the most important classes of error-correcting codes in [11]. In recent years, linear and cyclic over rings have been extensively studied, for example, the structure of $(1+u)$-constacyclic codes of arbitrary lengths over $\mathbb{F}_{2}+u\mathbb{F}_{2}$ have been studied and Gray images in [3,8]. The dual codes of $\mathbb{Z}_{2}$-double cyclic codes are studied in[9]. The finite chain rings of cyclic self-dual codes have been discussed in [6]. The Hamming distances of all binary codes of length $2^e$ have been determined in [10].

 Delsarte made a remarkable success about additive codes in 1973 in [1], he defined additive codes as subgroups of the underlying abelian group in a translation association scheme. A class of codes which contains all binary and quaternary codes as a subclass is called $\mathbb{Z}_{2}\mathbb{Z}_{4}$-additive codes. This class of codes have been studied in [4,5,7]. $\mathbb{Z}_{2}\mathbb{Z}_{2}[u]$-additive codes have been studied in [2].

In this Paper, we focus on $\mathbb{Z}_{2}\mathbb{Z}_{2}[u]$-$(1+u)$-additive constacyclic code. We also determine the form of the generators of $\mathbb{Z}_{2}\mathbb{Z}_{2}[u]$-$(1+u)$-additive constscyclic codes of arbitrary lengths and the parameters in terms of the degrees of the  generator polynomials of the code. We also study the generator polynomials of the dual code of a $\mathbb{Z}_{2}\mathbb{Z}_{2}[u]$-$(1+u)$-additive constacyclic code in terms of the generator polynomials of the Gray map of the code $C$.

\dse{2~~Preliminaries}
Since any $ \mathbb{Z}_{2}\mathbb{Z}_{2}[u]$-additive code $C$ must be closed under addition, it has to be closed under multiplication by elements in $R$ as well where $R= \mathbb{Z}_{2}+u \mathbb{Z}_{2}$, let $c=(a_0,a_1,\ldots,a_{\alpha-1},b_0,b_1,\ldots,\\
b_{\beta-1})\in C$ and $r=r_0+ur_1\in R$, we have
$$rc=(r_0a_0,r_0a_1,\ldots,r_0a_{\alpha-1},rb_0,rb_1,\ldots,rb_{ \beta-1}) \in C$$,
where $a_i\in \mathbb{Z}$, $0\leq i\leq \alpha-1$ and $b_j\in \mathbb{Z}_{2}+u\mathbb{Z}_{2}$, $0\leq j\leq \beta-1$.

In [2], an extension of the usual Gray map is defined as
$$ \phi: \mathbb{Z}_{2}^\alpha \times R^\beta \rightarrow \mathbb{Z}_{2}^n$$
$$ \phi(x,y)=(x,\phi(y_0),\phi(y_1),\ldots,\phi(y_{\beta-1})).$$
For any $x=(x_0,x_1,\ldots,x_{\alpha-1})\in\mathbb{Z}_{2}^\alpha$, $y=(y_0,y_1,\ldots,y_{\beta-1})\in R^ \beta$, where $n=\alpha+2\beta$ and  $\phi(0)=(0,0)$, $\phi(1)=(0,1)$, $\phi(u)=(1,1)$, $\phi(1+u)=(1,0)$.

The map $\phi$ is an isometry which transforms the Lee distance in $\mathbb{Z}_{2}^\alpha\times R^\beta$
to the Hamming distance in $\mathbb{Z}_{2}^n$, where $n=\alpha+2\beta$.

An inner product for two elements $c_1,c_2\in\mathbb{Z}_{2}^\alpha\times R^\beta$ is defined as
$$\langle c_1,c_2\rangle=u\sum_{i=0}^{\alpha-1}a_id_i+\sum_{j=0}^{\beta-1}b_je_j,$$
where $c_1=(a_0,a_1,\cdots,a_{\alpha-1},b_0,b_1,\cdots,b_{\beta-1})$, $c_2=(d_0,d_1,\cdots,d_{\alpha-1},e_0,e_1,\cdots,e_{\beta-1})$.

Let $C$ be a $\mathbb{Z}_{2}\mathbb{Z}_{2}[u]$-additive code. The dual of $C$, denoted by
$$ C^{\bot} =\{ a \in \mathbb{Z}_{2}^{\alpha} \times R^{\beta}| \langle a,b \rangle =0, \forall b \in C\}.$$

Let $C_X$(respectively $C_Y$) be the punctured code of $C$ by deleting the coordinates out of  $X$(respectively $Y$).
Let $C_b$ be the subcode of $C$ which contains all codewords as follow  $c=(x,uy)$, where $k_0$ is the dimension of $C_b$.\\

\noindent\textbf{Definition 2.1.} Let $C\subseteq\mathbb{Z}_{2}^\alpha\times R^\beta$ be a $\mathbb{Z}_{2}^\alpha\times R^\beta$-additive code, group isomorphic to $\mathbb{Z}_{2}^{k_1}\times R^{k_2}$, then $C$ is called a $\mathbb{Z}_{2}\mathbb{Z}_{2}[u]$-additive code of $Type(\alpha,\beta;k_1,k_2,k_0)$, where $k_0$ is defined above.\\

Let $k_0'$ and $k_2'$ be the dimensions of subcodes ${(a_0,a_1,\cdots,a_{\alpha-1},0,0,\cdots,0)\in C}$ and $(0,0,\cdots,0,\\
b_0,b_1,\cdots,b_{\beta-1})\in C-C_b$, respectively. Let $k_0''=k_0-k_0'$ and $k_2''=k_2-k_2'$, respectively.

Let $C$ be a $\mathbb{Z}_{2}\mathbb{Z}_{2}[u]$-additive code of $ Type( \alpha, \beta;k_1,k_2,k_0)$. Then,\\
$$|C|=2^{k_1}4^{k_2},\ |C^\bot|=2^{\alpha+k_1-2k_0}4^{\beta-k_1-k_2+k_0},$$
$$|C_X|=2^{k_0+k_2''},\ |C_X^\bot|=2^{\alpha-k_0-k_2''},$$
$$|C_Y|=2^{k_1-k_0'}4^{k_2},\ |C_Y^\bot|=2^{k_1-k_0'}4^{ \beta-k_1-k_2+k_0'},$$ where $k_0''=k_0-k_0'$, $k_2''=k_2-k_2'$.

\dse{3~~ $\mathbb{Z}_{2}\mathbb{Z}_{2}[u]$-$(1+u)$-additive constacyclic code}

Constacyclic codes usually are identified as ideals in a certain ring. In this section we show that $\mathbb{Z}_{2}\mathbb{Z}_{2}[u]$-$(1+u)$-additive constacyclic code of length $(\alpha,\beta)$ can be identified as $R[x]$-submodules of $R_{\alpha,\beta}=\mathbb{Z}_{2}[x]/\langle x^\alpha-1\rangle\times R[x]/\langle x^\beta-1-u\rangle$.\\

\noindent\textbf{Definition 3.1.} {A subset $C$ of $\mathbb{Z}_{2}^\alpha\times R^\beta$ is called a $\mathbb{Z}_{2}\mathbb{Z}_{2}[u]$-(1+u)-additive constacyclic code if
\\(1) $C$ is an additive code, and
\\(2) for any codeword $c=(a_0,a_1, \ldots ,a_{\alpha-1},b_0,b_1,\ldots,b_{\beta-1})\in C$. Its (1+u)-constacyclic shift $$ \varphi(c)=(a_{\alpha-1},a_0,\ldots,a_{\alpha-2},(1+u)b_{\beta-1},b_0,\ldots,b_{\beta-2})\in C.$$

A element $c=(a_0,a_1,\ldots,a_{\alpha-1},b_0,b_1,\ldots,b_{\beta-1})\in C$ can be identified with a module element consisting of two polynomials
$$c=(a_0+a_1x+\ldots+a_{\alpha-1}x^{\alpha-1},b_0+b_1x\ldots+b_{\beta-1}x^{\beta-1})=(a(x),b(x)),$$
in $R_{\alpha,\beta}$. This identification gives a one-to-one correspondence between the elements in $\mathbb{Z}_{2}^\alpha\times R^\beta$ and $R_{\alpha,\beta}$.}\\

Let $d(x)\in R[x]$, $(a(x),b(x))\in R_{\alpha,\beta}$ and consider the following multiplication operation
$$d(x)\ast(a(x),b(x))=(\bar{d}(x)a(x),d(x)b(x))$$
where $\bar{d}(x)$ is the reduction $\bar{d}(x)\equiv d(x)~mod~u$.\\

The multiplication operation on $R_{\alpha,\beta}$ lead to the following easily proven the Corollary.\\

\noindent\textbf{Corollary 3.2.} {The multiplication above is well-defined. Moreover, $R_{\alpha,\beta}$ is $R[x]$-module with respect to this multiplication.}\\

As is common in the discussion of cyclic and constacyclic codes, we can regard codewords of a cyclic and constacyclic codes $C$ as vectors or as polynomials interchangeably.\\

\noindent\textbf{Definition 3.3.} {A subset $C\subseteq R_{\alpha,\beta}$ is called a $\mathbb{Z}_{2}\mathbb{Z}_{2}[u]$-$(1+u)$-additive constacyclic code if and only if $C$ is a subgroup of $C\subseteq R_{\alpha,\beta}$, and if for all
$$c(x)=(a_0+a_1x+\cdots+a_{\alpha-1}x^{\alpha-1},b_0+b_1x+ \cdots +b_{\beta-1}x^{\beta-1})\in C$$
We have
$$x\ast c(x)=(a_{\alpha-1}+a_0x+ \cdots+a_{\alpha-2}x^{\alpha-1},(1+u)b_{\beta-1}+b_0x+\cdots+b_{\beta-2}x^{\beta-1})\in C$$}\\

\noindent\textbf{Definition 3.4.} {A code $C$ is a $\mathbb{Z}_{2}\mathbb{Z}_{2}[u]$-$(1+u)$-additive constacyclic code if and only if $C$ is a $R[x]$-submodules of $R_{\alpha,\beta}$.}\\

\noindent\textbf{Proposion 3.5.} {Let $\beta$ be odd integer and let $\mu$ be the map $\mu$: $(\mathbb{Z}_{2}[x]/\langle x^{\alpha}-1\rangle,R[x]/\langle x^{\beta}-1)\rangle\rightarrow (\mathbb{Z}_{2}[x]/\langle x^{\alpha}-1\rangle,R[x]/\langle x^{\beta}-1-u\rangle$, i.e $\mu(a(x),b(x))=(a(x),b((1+u)x))$, then we have $\mu$ is a ring isomorphism.}\\

\noindent\textbf{Theorem 3.6.} {Let $C\subseteq R_{\alpha,\beta}$ and $C_2(t)$ be the number of 2-cyclotomic classes modulo $t$ with $\alpha$, $\beta$ odd. Then the number of $\mathbb{Z}_{2}\mathbb{Z}_{2}[u]$-$(1+u)$-constacyclic codes is equal $2^{C_2(\alpha)}\times3^{C_2(\beta)}$.}\\

\noindent\textbf{Proof:} By [6, Theorem 5.1], the result is clearly.\\

The next contents gives the structure of $\mathbb{Z}_{2}\mathbb{Z}_{2}[u]$-$(1+u)$-additive in terms of the generator polynomials constacyclic code and the minimal generation sets.\\

\noindent\textbf{Lemma 3.7.}[3] {Let $C=\langle g(x)+up(x),ua(x)\rangle$ be any $(1+u)$-constacyclic code in $R[x]/(x^\beta-1)$, where $\beta=2^em$ and $gcd(2,m)=1$. Then $C$ must be one of the following forms:
\\ 1) $C=(g(x))$ where $g(x)|(x^\beta-1)~mod~2$;
\\ 2) $C=(ug(x))$ where $g(x)|(x^\beta-1)~mod~2$;
\\ 3) $C=(f_1^{i_1}(x)f_2^{i_2}(x)\cdots f_r^{i_r}(x))$ where $f_i(x)|(x^\beta-1)$ and for some $i_j$ we have $2^e\leq i_j\leq 2^{e+1}$.}\\

In [2], the map is defined as
 $$\sigma: C\rightarrow R[x]/\langle x^\beta-1-u\rangle,$$ where $\sigma(a(x),b(x))=b(x)$.
It is clear that $\sigma$ is module homomorphism whose image is a $R[x]$-submodule  of $R[x]/\langle x^\beta-1-u\rangle$ and $ker(\sigma)$ is a submodule of $C$, respctively. Moreover, $Im(\sigma)$ can easily be seen as an ideal of $R[x]/\langle x^\beta-1-u\rangle$.\\

Case1 $C=\langle g(x)\rangle$ where $g(x)|(x^\beta-1)~mod~2$.\\

 Note that
$$ ker(\sigma)=\{(a(x),0)\in C| a(x)\in Z_2[x]/\langle x^\alpha-1\rangle\}.$$

Define the set $I$ to be
$$ \{a(x)\in Z_2[x]/\langle x^\alpha-1\rangle|(a(x),0)\in ker(\sigma)\}.$$

It is clear that $I$ is an ideal and hence a cyclic code over $\mathbb{Z}_{2}[x]/\langle x^\alpha-1\rangle$,
$I=\langle a(x)\rangle$. By the first isomorphism theorem we have
$$C/ker(\sigma)\cong\langle g(x)\rangle.$$

Let $(l(x),g(x))\in C$ such that
$$\sigma(l(x),g(x))\in \langle g(x)\rangle.$$

This discussion shows that any $\mathbb{Z}_{2}\mathbb{Z}_{2}[u]$-$(1+u)$-additive constacyclic code can be generated as a $R[x]$-submodule of $R_{\alpha, \beta}$ by two elements of form $(a(x),0)$ and $(l(x),g(x))$, i.e. any element in the code $C$ can be described as
$$ h_1(x)\ast(a(x),0)+h_2(x)\ast(l(x),g(x)),$$
where $h_1(x)$, $h_2(x)\in R[x]$. We have $C=\langle(a(x),0),(l(x),g(x))\rangle$, where $a(x)$, $l(x)\in Z_2[x]$, $a(x)|(x^\alpha-1)$, $g(x)|(x^\beta-1)$.\\

Note that if $C=\langle(a(x),0),(l(x),g(x))\rangle$ is a $\mathbb{Z}_{2}\mathbb{Z}_{2}[u]$-$(1+u)$-additive constacyclic code, then the canonical projection $C_X$ is a cyclic code over $\mathbb{Z}_{2}$ generated by $\langle gcd(a(x),l(x))\rangle$ and $C_Y$ is a $(1+u)$-constacyclic code over $\mathbb{Z}_{2}+u\mathbb{Z}_{2}$ generated by $\langle g(x)\rangle$.\\

\noindent\textbf{Corollary 3.8.} {If $C=\langle(a(x),0),(l(x),g(x))\rangle$ is a $\mathbb{Z}_{2}\mathbb{Z}_{2}[u]$-$(1+u)$-additive constacyclic code, then we have $deg(l(x))<deg(a(x))$.}\\

\noindent\textbf{Proof:} The proof is similar to [4, Lemma 9].\\

\noindent\textbf{Lemma 3.9.} {If $C=\langle(a(x),0),(l(x),g(x))\rangle$ is a $\mathbb{Z}_{2}\mathbb{Z}_{2}[u]$-$(1+u)$-additive constacyclic code, then we have $a(x)|(x^{\beta}-1)l(x)$.}\\

\noindent\textbf{Proof:} Consider\\
$$\sigma((x^{\beta}-1-u)(l(x),g(x)))=\sigma(((x^{\beta}-1)l(x),0))\in C$$
Hence,\\
$((x^{\beta}-1)l(x),0)\in ker(\sigma)\subseteq C$ and $a(x)|(x^{\beta}-1)l(x)$.\\

\noindent\textbf{Theorem 3.10.} {Let $C=\langle(a(x),0),(l(x),g(x))\rangle$ is a $\mathbb{Z}_{2}\mathbb{Z}_{2}[u]$-$(1+u)$-additive constacyclic code in $R_{\alpha, \beta}$, $deg(a(x))=t_1$, $deg(g(x))=t_2$, $h(x)=(x^\beta-1)/g(x)$. Let\\
$$S_1=\bigcup_{i=0} ^{\alpha-t_1-1}\{x^i\ast(a(x),0)\},$$
$$S_2=\bigcup_{i=0} ^{\beta-t_2-1}\{x^i\ast(l(x),g(x))\},$$
$$S_3=\bigcup_{i=0} ^{t_2-1}\{x^i\ast(l(x)h(x),u)\}.$$
Then,  $$S=S_1 \bigcup S_2 \bigcup S_3$$
forms a minimal spanning set for $C$ as $R$-module. Moreover, $C$ has $2^{\alpha-t_1}4^{\beta-t_2}2^{t2}$ codewords.}\\

\noindent\textbf{Proof:} Let $c(x)$ be a codeword in $C$. Then there are polynomials $d_1(x)$, $d_2(x)\in R[x]$ such that \\
$$c(x)=\bar{d_1}(x)\ast(a(x),0)+d_2(x)\ast(l(x),g(x))$$
If $deg(\bar{d}_1(x))\leq\alpha-t_1-1$, then $\bar{d}_1(x)\ast(a(x),0)\in Span(S_1)$. Otherwise, by the division algorithm we get two polynomials $q_1(x)$, $r_1(x)\in R[x]$ such that $\bar{d}_1(x)=((\frac{x^\alpha-1}{a(x)}\bar{q}_1(x)+\bar{r}_1(x))$, where $\bar{r}_1(x)=0$ or $deg\bar{r}_1(x)\leq r-t_1-1$.\\ Hence,\\
$\bar{d}_1(x)\ast(a(x),0)=(\frac{x^\alpha-1}{a(x)}\bar{q}_1(x)+\bar{r}_1(x))\ast(a(x),0)=\bar{r}_1(x)\ast(a(x),0)$.
Therefore, we have $\bar{d}_1(x)\ast(a(x),0)\in Span(S_1)$.\\
If $deg(d_2(x))\leq\beta-t_2-1$, then $d_2(x)\ast(l(x),g(x))\in Span(S_2)$. Otherwise by the division algorithm again $d_2(x)=h(x)q_2(x)+r_2(x)$, where $r_2(x)=0$ or $deg(r_2(x))\leq\beta-t_2-1$. This implies that
$$d_2(x)\ast(l(x),g(x))=(h(x)q_2(x)+r_2(x))\ast(l(x),g(x))$$
$$=(l(x)h(x),g(x)h(x))\ast q_2(x)+r_2(x)\ast(l(x),g(x))$$
Note that since $r_2(x)=0$ or $deg(r_2(x))\leq\beta-t_2-1$, $r_2(x)\ast(l(x),g(x))\in Span(S_2)$. So we only need to show that $q_2(x)\ast(l(x)h(x),u) \in Span(S)$. If $deg(q_2(x))\leq t_2-1$, then we have done. Otherwise $q_2(x)=g(x)q_3(x)+r_3(x)$ where $r_3(x)=0$ or $deg(r_2(x))\leq t_2-1$. This implies that
$ q_2(x)\ast(l(x)h(x),u)=(g(x)q_3(x)+r_3(x))\ast(l(x)h(x),u)
=q_3(x)\ast(0,ug(x)) +r_3(x)\ast(h(x)l(x),u)$
Since $r_3(x)\ast(h(x)l(x),u)\in Span(S_3)$, we just need to show $((x^\beta-1)l(x),ug(x))\ast q_3(x)\in Span(S)$. Since $x^i\ast(0,ug(x))\in Span(S_2)$, $0\leq i\leq \beta-t_2-1$, we have $q_3(x)\ast(0,ug(x))\in Span(S_2)$. Therefore, we just need to show $q_3(x)\ast((x^\beta-1)l(x),0) \in Span(S)$. By Corollary 3.9, we have $a(x)|(x^{\beta}-1)l(x)$ and hence $ q_3(x)\ast((x^\beta-1)l(x),0)\in Span(S_1)$. Therefore, $S$ is a spanning set for $C$. It is also obvious that the $C$ is minimal in the sense that no element in $C$ is a linear combination of the other elements. Note also that $S_1$ will contribute $2^{\alpha-t_1}$ codewords, $S_2$ will contribute $4^{\beta-t_2}$, $S_3$ will contribute $2^{t_2}$ because $l(x)$ and $a(x)$ can be assumed to be binary polynomials.\\

Cast2  $c=\langle ug(x)\rangle$ where $g(x)|(x^\beta-1)~mod~2$\\

Note that if $C=\langle(a(x),0),(l(x),ug(x))\rangle$ is a $\mathbb{Z}_{2}\mathbb{Z}_{2}[u]$-$(1+u)$-additive constacyclic code, then the canonical projection $C_X$ and $C_Y$ are a cyclic code over $\mathbb{Z}_{2}$ and a $(1+u)$-constacyclic code over $\mathbb{Z}_{2}+u\mathbb{Z}_{2}$ generated by $\langle gcd(a(x),l(x))\rangle$ and $\langle ug(x)\rangle$.\\

\noindent\textbf{Lemma 3.11.} {If $C=\langle(a(x),0),(l(x),ug(x))\rangle$ is a $\mathbb{Z}_{2}\mathbb{Z}_{2}[u]$-$(1+u)$-additive constacyclic code, then we have $deg(l(x))<deg(a(x))$.}\\

\noindent\textbf{Corollary 3.12.} {If $C=\langle(a(x),0),(l(x),ug(x))\rangle$ is a $\mathbb{Z}_{2}\mathbb{Z}_{2}[u]$-$(1+u)$-additive constacyclic code, then we have $a(x)|(\frac{x^{\beta}-1}{g(x)}l(x)$.}\\

\noindent\textbf{Theorem 3.13.}  {Let $C=\langle(a(x),0),(l(x),ug(x))\rangle$ is a $\mathbb{Z}_{2}\mathbb{Z}_{2}[u]$-$(1+u)$-additive constacyclic code in $R_{\alpha, \beta}$, $deg(a(x))=t_1$,       $degg(x)=t_2$, $h(x)=(x^\beta-1)/g(x)$. Let
$$S_1=\bigcup_{i=0} ^{\alpha-t_1-1}\{x^i\ast(a(x),0)\},$$
$$S_2=\bigcup_{i=0} ^{\beta-t_2-1}\{x^i\ast(l(x),ug(x))\}.$$
Then, $$S=S_1 \bigcup S_2 $$
forms a minimal spanning set for $c$ as $R$-module. Moreover, $C$ has $2^{\alpha-t_1}2^{\beta-t_2}$ codewords.}\\

\noindent\textbf{Proof:} The proof is similar to Theorem 3.10.\\

Cast3 $C=\langle f_1^{i_1}(x)f_2^{i_2}(x)\cdots f_r^{i_r}(x)\rangle$ where $f_i(x)|(x^\beta-1)$ and for some $i_j$ we have $2^e\leq i_j\leq 2^{e+1}$.\\

Let $C=\langle f(x)g(x)\rangle$ where $g(x)$ is a polynomial of largest degree such that $deg(g(x))=t_2$, $degf(x)=t_3$ and $f(x)|g(x)|(x^\beta-1)~mod~2$.\\

Note that if $C=\langle(a(x),0),(l(x),f(x)g(x))\rangle$ is a $\mathbb{Z}_{2}\mathbb{Z}_{2}[u]$-$(1+u)$-additive constacyclic code, then the canonical projection $C_X$ and $C_Y$ are a cyclic code over $\mathbb{Z}_{2}$ and a $(1+u)$-constacyclic code over $\mathbb{Z}_{2}+u\mathbb{Z}_{2}$ generated by $\langle gcd(a(x),l(x))\rangle$ and $\langle f(x)g(x)\rangle$.\\

\noindent\textbf{Theorem 3.14.} {Let $C=\langle(a(x),0),(l(x),f(x)g(x))\rangle$ be a $\mathbb{Z}_{2}\mathbb{Z}_{2}[u]$-$(1+u)$-additive constacyclic code in $R_{\alpha, \beta}$, $dega(x)=t_1$, $degg(x)=t_2$, $degf(x)=t_3$, $f(x)|g(x)|(x^\beta-1)$, $h(x)=(x^\beta-1)/g(x)$. Let\\
$$S_1=\bigcup_{i=0} ^{\alpha-t_1-1}\{x^i\ast(a(x),0)\},$$
$$S_2=\bigcup_{i=0} ^{\beta-t_2-1}\{x^i\ast(l(x),f(x)g(x))\},$$
$$S_3=\bigcup_{i=0} ^{t_2-t_3-1}\{x^i\ast(l(x)h(x),uf(x))\}.$$
Then,  $$S=S_1 \bigcup S_2 \bigcup S_3$$
forms a minimal spanning set for $c$ as $R$-module. Moreover, $C$ has $2^{\alpha-t_1}4^{\beta-t_2}2^{t_2-t_3}$ codewords.}\\

\noindent\textbf{Proof:} Let $c(x)$ be a codeword in $C$. Then there are polynomials $d_1(x)$, $d_2(x)\in R[x]$ such that \\
$$c(x)=\bar{d_1}(x)\ast(a(x),0)+d_2(x)\ast(l(x),f(x)g(x))$$
If $deg(\bar{d}_1(x))\leq\alpha-t_1-1$, then $\bar{d}_1(x)\ast(a(x),0)\in Span(S_1)$. Otherwise, by the division algorithm we get two polynomials $q_1(x)$, $r_1(x)\in R[x]$ such that $\bar{d}_1(x)=(\frac{x^\alpha-1}{a(x)}\bar{q}_1(x)+\bar{r}_1(x))$, where $\bar{r}_1(x)=0$ or $deg\bar{r}_1(x)\leq r-t_1-1$. Hence,\\
$\bar{d}_1(x)\ast(a(x),0)=((\frac{x^\alpha-1}{a(x)})\bar{q}_1(x)+\bar{r}_1(x))\ast(a(x),0)=\bar{r}_1(x)\ast(a(x),0)$.
Therefore, we have $\bar{d}_1(x)\ast(a(x),0)\in Span(S_1)$.\\
If $deg(d_2(x))\leq\beta-t_2-1$, then $d_2(x)\ast(l(x),f(x)g(x))\in Span(S_2)$. Otherwise by the division algorithm again $d_2(x)=h(x)q_2(x)+r_2(x)$, where $r_2(x)=0$ or $deg(r_2(x))\leq\beta-t_2-1$. This implies that
$$d_2(x)\ast(l(x),f(x)g(x))=(h(x)q_2(x)+r_2(x))\ast(l(x),f(x)g(x))$$
$$=q_2(x)\ast(l(x)h(x),uf(x)) +r_2(x)\ast(l(x),f(x)g(x))$$
Note that since $r_2(x)=0$ or $degr_2(x)\leq\beta-t_2-1$, $r_2(x)\ast(l(x),g(x))\in Span(S_2)$. So we only need to show that $q_2(x)\ast(l(x)h(x),uf(x)) =(l(x),f(x)g(x))\ast\frac{x^\beta-1}{g(x)}q_2(x)\in Span(S)$. If $deg(q_2(x))\leq t_2-t_3-1$, then we have done. Otherwise by the division algorithm, we have $q_2(x)=\frac{g(x)}{f(x)}q_3(x)+r_3(x)$, where $r_3(x)=0$ or $degr_3(x)\leq t_2-t_3-1$. This implies that\\
$$q_2(x)\ast(l(x)h(x),ug(x))=(\frac{g(x)}{f(x)}q_3(x)+r_3(x))\ast(l(x)h(x),ug(x))$$
$$=\frac{g(x)}{f(x)}q_3(x)\ast(l(x)h(x),ug(x))+r_3(x)\ast(l(x)h(x),ug(x)).$$
Note that since $r_3(x)=0$ or $degr_3(x)\leq t_2-t_3-1$, $r_3(x))\ast(l(x)h(x),ug(x))\in Span(S_3)$. So we only need to show that $(\frac{g(x)}{f(x)}q_3(x))\ast(l(x)h(x),ug(x))=\frac{x^\beta-1}{f(x)}q_3(x)\ast(l(x),f(x)g(x))\in Span(S)$, If $deg(q_3(x))\leq t_2-t_3-1$, then we have done. Otherwise by the division algorithm, we have
$\frac{x^\beta-1}{f(x)}q_3(x)=\frac{x^\beta-1}{g(x)}q_4(x)+r_4(x)$ where $r_4(x)=0$ or $deg(r_4(x))\leq \beta-t_2-1$, $deg(q_4(x))\leq deg(q_3(x))$. This implies that
$$\frac{x^\beta-1}{f(x)}q_3(x)\ast(l(x),f(x)g(x))=(\frac{x^\beta-1}{g(x)}q_4(x)+r_4(x))\ast(l(x),f(x)g(x))$$
$$\frac{x^\beta-1}{g(x)}q_4(x)\ast(l(x),f(x)g(x))+r_4(x)\ast(l(x),f(x)g(x))$$
Note that since $r_4(x)=0$ or $deg(r_3(x))\leq\beta-t_2-1$, $r_4(x)\ast(l(x),g(x))\in Span(S_2)$. So we only need to show that $q_4(x)\ast(l(x)h(x),uf(x)) =\frac{x^\beta-1}{g(x)}q_4(x)\ast(l(x),f(x)g(x))\in Span(S)$. Just do it, we can always get $\frac{x^\beta-1}{f(x)}q_i(x)$, where $deg(q_i(x))\leq t_2-t_3-1$, and $q_i(x)\ast(l(x)h(x),uf(x)) =(l(x),f(x)g(x))\ast\frac{x^\beta-1}{g(x)}q_i(x)\in Span(S_2)$. Hence, we can obtain the result.\\

In the following contents, by using these spanning sets above, we can obtain the parameters $(\alpha,\beta;k_{1},k_{2},k_{0})$
of the code.\\

\noindent\textbf{Lemma 3.15.} {Let $C=\langle(a(x),0),(l(x),g(x))\rangle$ is a $\mathbb{Z}_{2}\mathbb{Z}_{2}[u]$-$(1+u)$-additive constacyclic code. Then, $$C_b=\langle(a(x),0),(l(x)h(x),u),(0,ug(x))\rangle$$}\\

\noindent\textbf{Proof:}  The result follows from the definition of $C_b$.\\

Let $C=\langle(a(x),0),(l(x),ug(x))\rangle$ is a $\mathbb{Z}_{2}\mathbb{Z}_{2}[u]$-$(1+u)$-additive constacyclic code, then we have $C_b=\langle(a(x),0),(l(x),ug(x))\rangle$. Let $C=\langle(a(x),0),(l(x),f(x)g(x))\rangle$, then we have $C_b=\langle(a(x),0),(l(x)\frac{x^\beta-1}{g(x)},uf(x)),(0,uf(x)g(x))\rangle$.\\

\noindent\textbf{Theorem 3.16.} {Let $C=\langle(a(x),0),(l(x),g(x))\rangle$ be a $\mathbb{Z}_{2}\mathbb{Z}_{2}[u]$-$(1+u)$-additive constacyclic code of $Type(\alpha,\beta;k_{11},k_{21},k_{01})$, where $g(x)h(x)=x^\beta-1$. Then
$$k_{01}=\alpha-deg(gcd(l(x)h(x),a(x))),$$ $$k_{11}=\alpha+degg(x)-dega(x),$$ $$k_{21}=\beta-degg(x).$$}\\

\noindent\textbf{Proof:}  the parameters $k_{11}$ and $k_{21}$ are known from Theorem 3.10 and the parameters $k_{01}$ is the dimension of parameters $(C_b)_X$. By Lemma 3.16, the space $(C_b)_X$ is generated by the polynomials $a(x)$ and $l(x)h(x)$. Since the ring is a polynomials ring and thus a principal ideal ring, it is generated by the greatest common divisor of the two polynomials. Then, $k_{01}=\alpha-deg(gcd(l(x)h(x),a(x)))$.\\

By Theorem 3.16, we know the generator polynomials of $C_X$ and $(C_b)_X$, which are $gcd(a(x),l(x))$, and  $gcd(a(x),l(x)h(x))$, respectively. Hence, we have  $$k_{01}'=\alpha-dega(x),\ k_{01}''=deg(a(x))-deg(gcd(a(x),l(x)h(x))),$$ $$k_{21}'=\beta-deg(g(x))-k_2'',\ k_{21}''=deg(gcd(a(x),l(x)h(x)))-deg(gcd(a(x),l(x))).$$\\

\noindent\textbf{Theorem 3.17.}  {Let $C=\langle(a(x),0),(l(x),ug(x))\rangle$ be a $\mathbb{Z}_{2}\mathbb{Z}_{2}[u]$-$(1+u)$-additive constacyclic code of $Type(\alpha,\beta;k_{02},k_{12},k_{22})$. Then,
$$k_{02}=\alpha-deg(gcd(l(x),a(x))),$$ $$k_{12}=\alpha+\beta-deg(a(x))-deg(g(x)),$$ $$k_{22}=0.$$}\\

\noindent\textbf{Proof:}  The proof is similar to Theorem 3.16\\

By Theorem 3.17, we know the generator polynomials of $C_X$ and $(C_b)_X$, which are $gcd(a(x),l(x))$. Hence, we have
$$k_{02}'=\alpha-deg(a(x)),\ k_{02}''=deg(a(x))-deg(gcd(a(x),l(x))),\ k_{22}'=0,\ k_{22}''=0.$$\\

\noindent\textbf{Theorem 3.18.}  {Let $C=\langle(a(x),0),(l(x),f(x)g(x))\rangle$ is a $\mathbb{Z}_{2}\mathbb{Z}_{2}[u]$-$(1+u)$-additive constacyclic code of $Type(\alpha,\beta;k_{03},k_{13},k_{23})$. Then,
$$k_{03}=\alpha-deg(gcd(l(x),a(x))),$$ $$k_{13}=\alpha+deg(g(x))-deg(a(x))-deg(f(x)),$$ $$k_{23}=\beta-deg(g(x)).$$}\\

\noindent\textbf{Proof:}  The proof is similar to Theorem 3.16.\\

By Theorem 3.18, we know the generator polynomials of $C_X$ and $(C_b)_X$, which are $gcd(a(x),l(x))$, $gcd(a(x),l(x)h(x))$. Hence, we have
$$k_{03}'=\alpha-deg(a(x)),\ k_{03}''=deg(a(x))-deg(gcd(a(x),l(x))),$$ $$k_{32}'=\beta-deg(g(x))-k_{23}'',\ k_{23}''=deg(gcd(a(x),l(x)))-deg(gcd(a(x),l(x)h(x))).$$\\

We introduce some examples for this family of codes with good parameters.\\

\noindent\textbf{Example 1}  Let $R_{2,3}=\mathbb{Z}_{2}[x]/(x^2-1)\times R[x]/(x^3-1-u)$, $C$ be a $\mathbb{Z}_{2}\mathbb{Z}_{2}[u]$-$(1+u)$-additive constacyclic code in the form of $C=(1+x,1+x)$. The Gray image gives the optimal binary code with parameters[8,6,2].\\

\noindent\textbf{Example 2}   Let $R_{2,3}=\mathbb{Z}_{2}[x]/(x^7-1)\times R[x]/(x^7-1-u)$, $C$ be a $\mathbb{Z}_{2}\mathbb{Z}_{2}[u]$-$(1+u)$-additive constacyclic code in the form of $C=(1+x+x^2+x^4,u(1+x))$. The Gray image gives the optimal binary code with parameters[21,6,8].\\

\noindent\textbf{Example 3}   Let $R_{2,3}=\mathbb{Z}_{2}[x]/(x^2-1)\times R[x]/(x^6-1-u)$, $C$ be a $\mathbb{Z}_{2}\mathbb{Z}_{2}[u]$-$(1+u)$-additive constacyclic code in the form of $C=(1+x,1+x+x^3+x^5)$. The Gray image gives the optimal binary code with parameters[14,7,4].

\dse{4~~The dual of $\mathbb{Z}_{2}\mathbb{Z}_{2}[u]$-$(1+u)$-additive constacyclic code}

 In this section we show that the dual of $\mathbb{Z}_{2}\mathbb{Z}_{2}[u]$-$(1+u)$-additive constacyclic code is also $(1+u)$-constacyclic. We also study the generator polynomials of the dual code $C^\bot$, where $C$ is a separable $\mathbb{Z}_{2}\mathbb{Z}_{2}[u]$-$(1+u)$-additive constacyclic code.\\

\noindent\textbf{Theorem 4.1.} {If $C$ is any $\mathbb{Z}_{2}\mathbb{Z}_{2}[u]$-$(1+u)$-additive constacyclic code, then $C^\bot$ is also a $(1+u)$-additive constacyclic.}\\

\noindent\textbf{Proof:}   Let $C$ be any $\mathbb{Z}_{2}\mathbb{Z}_{2}[u]$-$(1+u)$-additive constacyclic code. Suppose $c_1=(a_0,a_1£¬\cdots£¬a_{\alpha-1},b_0,\\
b_1,\cdots,b_{\beta-1})\in C^\bot$. It suffices to show $\delta(c_1)\in C^\bot$. For any codeword $c_2=(d_0,d_1,\cdots,d_{\alpha-1},e_0,\\
e_1,\cdots,e_{\beta-1})\in C$, we have
$$ \langle c_1,c_2\rangle=u\sum_{i=0}^{\alpha-1}a_id_i+\sum_{j=0}^{\beta-1}b_je_j=0$$
The proof will be completed if we show that $\langle \delta(c_1),c_2\rangle=0$. Let $m=2gcd(\alpha,\beta)$, then $\delta^m(c_2)=c_2$, where $c_2\in C$. Let
$$\omega=\delta^{m-1}(c_2)=(d_1,d_2,\cdots,d_0,e_1,e_2,\cdots,(1+u)e_0).$$
Since $C$ is a $\mathbb{Z}_{2}\mathbb{Z}_{2}[u]$-$(1+u)$-additive constacyclic code, $\omega\in C$. Hence,\\
$0=\langle c_1,\omega\rangle=u(a_0d_1+a_1d_2+\cdots+a_{\alpha-1}d_0)+(b_0e_1+b_1e_2+\cdots+(1+u)b_{\beta-1}e_0)$
$=u(a_{\alpha-1}d_0+a_1d_2+\cdots+a_0d_1)+((1+u)b_{\beta-1}e_0+b_1e_2+\cdots+b_0e_1)=\langle \delta(c_1),c_2\rangle$.
Therefore, $\delta(c_1)\in C^\bot$, and hence $C^\bot$ is a $(1+u)$-additive constacyclic code.\\

The dual code of $\mathbb{Z}_{2}\mathbb{Z}_{2}[u]$-$(1+u)$-additive constacyclic code is also a $\mathbb{Z}_{2}\mathbb{Z}_{2}[u]$-$(1+u)$-additive constacyclic code. So, we have
\\(1)  $C=\langle(a(x),0)(l(x),g(x))\rangle$, then $C^\bot=\langle(\bar{a}(x),0)(\bar{l}(x),u\bar{g}(x))\rangle$, where $\bar{a}(x)|(x^\alpha-1)$, $\bar{g}(x)|(x^\beta-1)\ mod\ 2$,
\\(2)  $C=\langle(a(x),0)(l(x),ug(x))\rangle$, then $C^\bot=\langle(\bar{a}(x),0)(\bar{l}(x),\bar{g}(x))\rangle$, where $\bar{a}(x)|(x^\alpha-1)$, $\bar{g}(x)|(x^\beta-1)\ mod\ 2$,
\\(3)  $C=\langle(a(x),0)(l(x),f(x)g(x))\rangle$, then $C^\bot=\langle(\bar{a}(x),0)(\bar{l}(x),\bar{f}(x)\bar{g}(x)\rangle$, where $\bar{a}(x)|(x^\alpha-1)$, $\bar{f}(x)|\bar{g}(x)|(x^\beta-1)\ mod\ 2$.\\

\noindent\textbf{Theorem 4.2.}  {Let $C=\langle(a(x),0),(l(x),g(x))\rangle$ be a $\mathbb{Z}_{2}\mathbb{Z}_{2}[u]$-$(1+u)$-additive constacyclic code of $Type(\alpha,\beta;k_{01},k_{11},k_{21})$, and the dual code is $C^\bot=\langle(\bar{a}(x),0)(\bar{l}(x),u\bar{g}(x))\rangle$ where $h(x)=\frac{x^\beta-1}{g(x)}$, $\bar{a}(x)|(x^\alpha-1)$, $\bar{g}(x)|(x^\beta-1)\ mod\ 2$. Then
$$deg(\bar{a}(x))=\alpha-deg(gcd(a(x),l(x)h(x))),$$ $$deg(\bar{g}(x))=deg(h(x))+deg(a(x))-deg(gcd(a(x),l(x)h(x))).$$}\\

\noindent\textbf{Proof:}  Let $C^\bot$ be a code of $Type(\alpha,\beta;\bar{k}_{01}\bar{k}_{11},\bar{k}_{21})$. It is easy to prove that $(C_X)^\bot$ is a binary cyclic code generated by $\bar{a}(x)$, so $|(C_X)^\bot|=2^{\alpha-deg\bar{a}(x)}$. Moreover, $|(C_X)^\bot|=2^{\alpha-k_{21}''-k_0}$, it is known that $\bar{k}_{01}=\alpha-k_{01}$, $\bar{k}_{11}=\alpha+k_{11}-2k_{01}$, $\bar{k}_{21}=\beta-k_{21}-k_{11}+k_{01}$. we can obtain the result.\\

\noindent\textbf{Corollary 4.3.}  {Let $C=\langle(a(x),0),(l(x),ug(x))\rangle$ be any $\mathbb{Z}_{2}\mathbb{Z}_{2}[u]$-$(1+u)$-additive constacyclic code, and the dual code $C^\bot=\langle(\bar{a}(x),0)(\bar{l}(x),\bar{g}(x))\rangle$ where $\bar{a}(x)|(x^\alpha-1)$, $\bar{g}(x)|(x^\beta-1)$. Then
$$deg(\bar{a}(x))=\alpha-deg(gcd(a(x),l(x))),$$
$$deg(\bar{g}(x))=deg(h(x))-deg(a(x))+deg(gcd(a(x),l(x))).$$}\\

\noindent\textbf{Corollary 4.4.}  {Let $C=\langle(a(x),0),(l(x),f(x)g(x))\rangle$ is a $\mathbb{Z}_{2}\mathbb{Z}_{2}[u]$-$(1+u)$-additive constacyclic code of $Type(\alpha,\beta;k_{03},k_{13},k_{23})$, and the dual code $C^\bot=\langle(\bar{a}(x),0)(\bar{l}(x),\bar{f}(x)\bar{g}(x)\rangle$,  where $\bar{a}(x)|(x^\alpha-1)$, $\bar{f}(x)|\bar{g}(x)|(x^\beta-1)\ mod\ 2$. Then
$$deg(\bar{a}(x))=\alpha-deg(gcd(a(x),l(x)h(x))),$$
$$deg(\bar{g}(x))=\beta-deg(f(x))-deg(a(x))-deg(gcd(a(x),l(x))),$$
$$deg(\bar{f}(x))=deg(h(x))+2deg(a(x))+deg(f(x))-deg(g(x))-2deg(gcd(a(x),l(x))).$$\\

We know that a $\mathbb{Z}_{2}\mathbb{Z}_{2}[u]$-additive code $C$ is separable if and only if $C^\bot$ is separable. Moreover, if a $\mathbb{Z}_{2}\mathbb{Z}_{2}[u]$-$(1+u)$-additive constacyclic code is separable, then it is easy to find the generator polynomials of the dual, which are given in the following theorems.\\

\noindent\textbf{Theorem 4.5.}  {Let $c=\langle(a(x),0),(0,g(x))\rangle$ be a $\mathbb{Z}_{2}\mathbb{Z}_{2}[u]$-$(1+u)$-additive constacyclic code. Then
$$C^\bot=\langle(\frac{x^\alpha-1}{a^\ast(x)},0),(0,u\frac{x^\beta-1}{g^\ast(x)})\rangle.$$}\\

\noindent\textbf{Theorem 4.6.}  {Let $c=\langle(a(x),0),(0,ug(x))\rangle$ be a $\mathbb{Z}_{2}\mathbb{Z}_{2}[u]$-$(1+u)$-additive constacyclic code. Then
$$C^\bot=\langle(\frac{x^\alpha-1}{a^\ast(x)},0),(0,\frac{x^\beta-1}{g^\ast(x)})\rangle.$$}\\

\noindent\textbf{Theorem 4.7.}  {Let $c=\langle(a(x),0),(0,f_1^{i_1}(x)f_2^{i_2}(x)\cdots f_r^{i_r}(x))\rangle$ be a $\mathbb{Z}_{2}\mathbb{Z}_{2}[u]$-$(1+u)$-additive constacyclic code. Then
$$C^\bot=\langle(\frac{x^\alpha-1}{a^\ast(x)},0),(0,(f_1^{2^{e+1}-i_1}(x))^\ast(f_2^{2^{e+1}-i_2}(x))^\ast\cdots(f_r^{2^{e+1}-i_r}(x))^\ast)\rangle.$$}\\

\dse{5~~Gary Images of $\mathbb{Z}_{2}\mathbb{Z}_{2}[u]$-$(1+u)$-additive constacyclic code}

In this section we obtain that the generators polynomials for the dual of $\mathbb{Z}_{2}\mathbb{Z}_{2}[u]$-$(1+u)$-additive constacyclic code by using the Gary images.\\

\noindent\textbf{Theorem 5.1.}  {Let $C$ be a $\mathbb{Z}_{2}\mathbb{Z}_{2}[u]$-$(1+u)$-additive constacyclic code of length $(\alpha,\beta)$, then $\phi(C)$ is a binary double cyclic code of length $(\alpha,2\beta)$.}\\

\noindent\textbf{Theorem 5.2.}  {Let $C=\langle(a(x),0),(l(x),g(x))\rangle$ be a $\mathbb{Z}_{2}\mathbb{Z}_{2}[u]$-$(1+u)$-additive constacyclic code of length $(\alpha,\beta)$, then $\phi(C)=\langle(a(x),0),(l(x),g(x))\rangle$ is a binary double cyclic code of length $(\alpha,2\beta)$.}\\

\noindent\textbf{Proof:}  Since the canonical projections $C_X$ and $C_X$ are a cyclic code over $\mathbb{Z}_{2}$ and a $(1+u)$-constacyclic code over $\mathbb{Z}_{2}+u\mathbb{Z}_{2}$ generated by $gcd(a(x),l(x))$ and $g(x)$, respectively. By[3, Lemma 11], we can obtain the result.\\

\noindent\textbf{Corollary 5.3.}  {Let $C=\langle(a(x),0),(l(x),ug(x))\rangle$ be a $\mathbb{Z}_{2}\mathbb{Z}_{2}[u]$-$(1+u)$-additive constacyclic code of length $(\alpha,\beta)$, then $\phi(C)=\langle(a(x),0),(l(x),g(x^\beta-1))\rangle$ is a binary double cyclic code of length $(\alpha,2\beta)$.}\\

\noindent\textbf{Corollary 5.4.}  {Let $C=\langle(a(x),0),(l(x),f_1^{i_1}(x)f_2^{i_2}(x)\cdots f_r^{i_r}(x))\rangle$ be a $\mathbb{Z}_{2}\mathbb{Z}_{2}[u]$-$(1+u)$-additive constacyclic code of length $(\alpha,\beta)$, then $\phi(C)=\langle(a(x),0),(l(x),f_1^{i_1}(x)f_2^{i_2}(x)\cdots f_r^{i_r}(x))\rangle$ is a binary double cyclic code of length $(\alpha,2\beta)$.}\\

\noindent\textbf{Corollary 5.5.} {Let $C=\langle(a(x),0),(l(x),g(x))\rangle$ be a $\mathbb{Z}_{2}\mathbb{Z}_{2}[u]$-$(1+u)$-additive constacyclic code of length $(\alpha,\beta)$, then $\phi(C)$ is a binary linear code of dimension $\alpha+2\beta-dega(x)-degg(x)$.}\\

Similarity, we can obtain if $C=\langle(a(x),0),(l(x),ug(x))\rangle$ be a $\mathbb{Z}_{2}\mathbb{Z}_{2}[u]$-$(1+u)$-additive constacyclic code, $\phi(C)$ is a binary linear code of dimension $\alpha+2\beta-dega(x)-degg(x^\beta-1)$; if $C=\langle(a(x),0),(l(x),f_1^{i_1}(x)f_2^{i_2}(x)\cdots f_r^{i_r}(x))\rangle$, $\phi(C)$ is a binary linear code of dimension of $\alpha+2\beta-dega(x)-\sum_{j=0}^{r}i_jdegf_j$.\\

In following contents, we will show some properties of $\mathbb{Z}_{2}\mathbb{Z}_{2}[u]$-$(1+u)$-additive constacyclic code and its dual.\\

\noindent\textbf{Theorem 5.6.}  {Let $C$ be a $\mathbb{Z}_{2}\mathbb{Z}_{2}[u]$-additive self-dual code of length $(\alpha,\beta)$, then $\phi(C)$ is a binary self-dual code of length $\alpha+2\beta$.}\\

\noindent\textbf{Proof:}  For any $c_1=(a_1,x_1+uy_1), c_2=(a_2,x_2+uy_2)\in C$ where $a_1,a_2\in \mathbb{Z}_{2}^\alpha$, $x_1+uy_1,x_2+uy_2\in R^\beta$. For $C$ is a $\mathbb{Z}_{2}\mathbb{Z}_{2}[u]$-$(1+u)$-additive self-dual code $c_1c_2=ua_1a_2+x_1x_2+u(x_1y_2+x_2y_1)=x_1x_2+u(x_1y_2+x_2y_1+a_1a_2)=0$, then $x_1x_2=x_1y_2+x_2y_1+a_1a_2=0$. We have
$\phi(c_1)\cdot\phi(c_2)=(a_1,y_1,x_1+y_1)\cdot(a_2,y_2,x_2+y_2)=x_1x_2+x_1y_2+x_2y_1+a_1a_2=0$, we have $\phi(C)$ is a binary self-dual code of length $\alpha+2\beta$.\\

\noindent\textbf{Corollary 5.7.}  {Let $C$ be a $\mathbb{Z}_{2}\mathbb{Z}_{2}[u]$-$(1+u)$-additive constacyclic self-dual code of length $(\alpha,\beta)$, then $\phi(C)$ is a binary double cyclic self-dual code of length $\alpha+2\beta$.}\\

\noindent\textbf{Lemma 5.8.}  {Let $C$ be a $\mathbb{Z}_{2}\mathbb{Z}_{2}[u]$-additive code of $Type(\alpha,\beta;k_0,k_1,k_2)$ of length $(\alpha,\beta)$,The dual of $C$ is denoted by
$ C^{\bot}$, then $\phi(C^\bot)=\phi(C)^\bot$ is a binary code of $\alpha+2\beta$.}\\

\noindent\textbf{Proof:}  For any $c_1=(a_1,x_1+uy_1)\in C$, $c_2=(a_2,x_2+uy_2)\in C^\bot$, then $c_1\cdot c_2=0$, so, $x_1x_2=x_1y_2+x_2y_1+a_1a_2=0$, thus we can obtain $\phi(c_1)\cdot\phi(c_2)=x_1x_2+x_1y_2+x_2y_1+a_1a_2=0$, which means $\phi(C^\bot)\subseteq\phi(C)^\bot$, since $\phi$ is bijection, we know $|\phi(C)|=|C|=2^{2k_2+k_0+k_1}$, so $|\phi(C)^\bot|=2^{\alpha+2\beta-2k_2-k_0-k_1}$ and $|\phi(C^\bot)|=|C^\bot|=2^{\alpha+2\beta}/|C|=2^{\alpha+2\beta-2k_2-k_0-k_1}$, therefore $\phi(C^\bot)=\phi(C)^\bot$.\\

In [6], we have the following map:
$$\eta:\mathbb{Z}_{2}/(x^\alpha-1)\times\mathbb{Z}_{2}/(x^\beta-1)\rightarrow\mathbb{Z}_{2}/(x^m-1),$$
where $m=2lcm(\alpha,\beta)$, such that for any $c_1(x)=(c_{11}(x),c_{12}(x))$ and $c_2(x)=(c_{21}(x),c_{22}(x))$ of $\mathbb{Z}_{2}/(x^\alpha-1)\times\mathbb{Z}_{2}/(x^\beta-1)$, we have $\eta(c_1(x),c_2(x))=c_{11}(x)\theta_{\frac{m}{\alpha}}(x^\alpha)x^{m-1-degc_{21}(x)}c_{21}^\ast(x)+c_{12}(x)\theta_{\frac{m}{\beta}}(x^\beta)x^{m-1-degc_{22}(x)}c_{22}^\ast(x)$, where we denote the polynomial $\sum_{i=0}^{t-1}x^i$ by $\theta_t(x)$. The map $\eta$ is a bilinear map between $\mathbb{Z}_{2}[x]$-modules.\\

\noindent\textbf{Lemma 5.9.} {Let $c_1(x)=(c_{11}(x),c_{12}(x))$ and $c_2(x)=(c_{21}(x),c_{22}(x))$ be elements of $\mathbb{Z}_{2}/(x^\alpha-1)\times\mathbb{Z}_{2}/(x^\beta-1)$ such that $\eta(c_1(x),c_2(x))=0$. If $c_{12}(x)$ or $c_{22}(x)$ equals to 0, then $c_{11}(x)c_{21}^\ast(x)=0\ mod\ (x^\alpha-1)$, If $c_{11}(x)$ or $c_{21}(x)$ equals to 0, then $c_{12}(x)c_{22}^\ast(x)=0\ mod\ (x^\beta-1)$.}\\

\noindent\textbf{Corollary 5.10.}   {Let $C=\langle(a(x),0),(l(x),g(x))\rangle$ be a $\mathbb{Z}_{2}\mathbb{Z}_{2}[u]$-$(1+u)$-additive constacyclic code of length $(\alpha,\beta)$ and $C^\bot=\langle(\bar{a}(x),0)(\bar{l}(x),u\bar{g}(x))\rangle$. Then
$$\bar{a}(x)=\frac{x^\alpha-1}{gcd(a(x),l(x))^\ast},$$
$$\bar{g}(x^\beta-1)=\frac{(x^\beta-1)gcd(a(x),l(x))^\ast}{a^\ast(x)g^\ast(x)},$$
$$\bar{l}(x)=\frac{(x^\alpha-1)}{a^\ast(x)}\lambda(x).$$
where $\lambda(x)\in \mathbb{Z}_{2}$.}\\

\noindent\textbf{Proof:}  By Theorem 11, $\phi(C)=\langle a(x),0),(l(x),g(x))\rangle$ and $\phi (C^\bot)=\langle(\bar{a}(x),0)(\bar{l}(x),\bar{g}(x^\beta-1))\rangle$, and by Proposition 4.12[6],  Proposition 4.13[6],  Proposition 4.16[6], we have $\phi(C)^\bot=((\frac{x^\alpha-1}{gcd(a(x),l(x))^\ast},0)(\frac{(x^\alpha-1)}{a^\ast(x)}\lambda(x),\frac{(x^{2\beta}-1)gcd(a(x),l(x))^\ast}{a^\ast(x)g^\ast(x)}))$   and $\phi(C^\bot)=((\bar{a}(x),0)(\bar{l}(x),\bar{g}(x^\beta-1)))$, where $\lambda(x)\in \mathbb{Z}_{2}$, by Lemma 3, $\phi(C^\bot)=\phi(C)^\bot$, we can obtain  the results.\\

\noindent\textbf{Corollary 5.11.}   {Let $C=\langle(a(x),0),(l(x),ug(x))\rangle$ be a $\mathbb{Z}_{2}\mathbb{Z}_{2}[u]$-$(1+u)$-additive constacyclic code of length $(\alpha,\beta)$ and $C^\bot=\langle(\bar{a}(x),0)(\bar{l}(x),\bar{g}(x))\rangle$. Then
$$\bar{a}(x)=\frac{x^\alpha-1}{gcd(a(x),l(x))^\ast},$$
$$\bar{g}(x^\beta-1)=\frac{(x^{2\beta}-1)gcd(a(x),l(x))^\ast}{a^\ast(x)g^\ast(x^\beta-1)},$$
$$\bar{l}(x)=\frac{(x^\alpha-1)}{a^\ast(x)}\lambda(x).$$
where $\lambda(x)\in \mathbb{Z}_{2}$.}\\

\noindent\textbf{Proof:}  The proof is similar to corollary 5.8.\\

\noindent\textbf{Corollary 5.12.}   {Let $C=\langle(a(x),0),(l(x),f_1^{i_1}(x)f_2^{i_2}(x)\cdots f_r^{i_r}(x))\rangle$ be a $\mathbb{Z}_{2}\mathbb{Z}_{2}[u]$-$(1+u)$-additive constacyclic code of length $(\alpha,\beta)$ and $C^\bot=\langle(\bar{a}(x),0)(\bar{l}(x),(\bar{f}_1(x))(\bar{f}_2(x))\cdots(\bar{f}_r(x)))\rangle$. Then
$$\bar{a}(x)=\frac{x^\alpha-1}{gcd(a(x),l(x))^\ast},$$
$$\bar{f}_j(x)=\frac{(x^{2\beta}-1)gcd(a(x),l(x))^\ast}{a^\ast(x)f^{\ast}_j(x)^{i_j}},$$
$$\bar{l}(x)=\frac{(x^\alpha-1)}{a^\ast(x)}\lambda(x).$$
where $\lambda(x)\in \mathbb{Z}_{2}$.}\\

\noindent\textbf{Proof:}  The proof is similar to corollary 5.8.\\

\dse{ Conclusion}

In this paper, we determine the generator polynomials of such
constacyclic codes over $\mathbb{Z}_{2}\mathbb{Z}_{2}[u]$ and give the the minimal generating sets. we also determine the relationship of generators between the $\mathbb{Z}_{2}\mathbb{Z}_{2}[u]$-(1+u) additive constacyclic codes and its dual and study between the Gray images and (1+u)-additive constacyclic codes. It would be interesting to consider other classes of constacyclic codes over $\mathbb{Z}_{2}\mathbb{Z}_{2}[u]$.

\end{document}